\documentclass[review]{elsarticle}

\usepackage{amsfonts,amssymb}
\usepackage{amsmath}
\usepackage{ntheorem}
\biboptions{sort&compress}
\newtheorem*{Proof}{Proof}
\newtheorem{Theorem}{Theorem}
\newtheorem{Lemma}{Lemma}

\newtheorem{Property}{Property}
\allowdisplaybreaks

\journal{Journal of \LaTeX\ Templates}

\begin{document}

\begin{frontmatter}

\title{ On the derivatives  of rational B\'{e}zier curves\tnoteref{mytitlenote}}

\author{Mao Shi\fnref{myfootnote}}
\address{School of Mathematics and Statistics of Shaanxi Normal University, Xi'an China}

\ead{shimao@snnu.edu.cn}


\begin{abstract}
By studying the existing higher order derivation formulas  of rational B\'{e}zier curves, we find that they fail when the order of the derivative exceeds the degree of the curves. In this paper, we present a new derivation formula for rational B\'{e}zier curves that overcomes this drawback and show that the $k$th degree derivative of a $n$th degree rational B\'{e}zier curve can be written in terms of a $(2^kn)$th degree rational B\'{e}zier curve.we also consider the properties of the endpoints and the bounds of the derivatives.
\end{abstract}


\end{frontmatter}


\section{Introduction}

Rational B\'{e}zier curves are essential mathematical tools in CAGD  and can be defined as {\cite{Farin2002}}
\begin{equation}\label{eq:001}
  \boldsymbol{c}\left( t \right) =\frac{\boldsymbol{r}(t)}{\omega(t)}=\frac{\sum\limits_{i=0}^n{\omega _i\boldsymbol{r}_iB_{i}^{n}\left( t \right)}}{\sum\limits_{i=0}^n{\omega _iB_{i}^{n}\left( t \right)}},\ t\in \left[0,1\right],
\end{equation}
where $B_i^n(t)= {n \choose i}t^i(1-t)^{n-i}$ are Bernstein basis functions of degree $n$, $\boldsymbol{r}_i\in\mathbb{R}^d$ are control points and $\omega_i$ are the corresponding positive weights.

When all   weights are equal and nonzero, the rational B\'{e}zier curve is reduced to an integer B\'{e}zier curve
  \begin{equation} \label{eq:002}
 \boldsymbol{p}(t)=\sum\limits_{i=0}^{n}\boldsymbol{p}_iB_i^n(t).
 \end{equation}

 Although the rational B\'{e}zier curves  share the  properties of B\'{e}zier curves, such as evaluation, subdivision, degree elevation and the convex hull property , they have their own non-uniformly convergent and divergent properties as the weight(s) $\omega_{i}$ tend to positive infinity {\cite{Shi2005}}{\cite{Shi2018}}. Furthermore,  unlike the $k$th  derivative of a B\'{e}zier curve, it can be expressed as a $(n-k)$th degree B\'{e}zier curve using the formula
\begin{equation}\label{eq:003}
\boldsymbol{p}^{(k)}(t)=\frac{n!}{(n-k)!}\sum_{j=0}^{n-k}\Delta^k\boldsymbol{p}_jB_{j}^{n-k}\left( t \right), \ (n \geq k)
\end{equation}
where $\triangle^k$ is the $k$th differenc, the derivative of rational B\'{e}zier curves is not so well expressed explicitly, so computing the  higher order derivative of a rational B\'{e}zier curve is a challenging problem {\cite{Sederberg1987}}{\cite{Floater1992}}.

Using the Leibniz rule, the $k$th derivative of a rational B\'{e}zier curve can be expressed as \cite{Farin2002}\cite{Hollig2013}
\begin{equation}\label{eq:001b}
  \boldsymbol{c}^{(k)}\left( t \right) =\frac{\boldsymbol{r}^{(k)}(t)-\sum_{j=1}^{k}\binom{k}{j}\omega^{(j)}(t)\boldsymbol{c}^{(k-j)}\left( t \right)}{\omega(t)},
\end{equation}
but  Wang and Wang\cite{wang1995} pointed out that equation \eqref{eq:001b} has  two weaknesses. First, the geometric meaning is not obvious, and second, the all lower order derivatives of $\boldsymbol{c}^{(s)}\left( t \right) \  (s=1,...,k-1)$ would not be necessary.  Therefore, they proposed a derivative formula that $\boldsymbol{c}^{(k)}\left( t \right)$ can be represented as a function of $\Delta\boldsymbol{r}_i$ and be calculated directly. The key formula  in  \cite{wang1995} is
 \[
\frac{{d^s \left( {E_i^{n - 1} } \right)}}{{dt^s }} = \frac{{{\sum\limits_{k = 0}^i {\sum\limits_{j = i + 1}^n {\sum\limits_{0 \le l_1 ,...,l_s  \le n}^{} {\omega _k \omega _j \omega _{l_1 }  \cdots \omega _{l_s } \left( {j - k} \right) \binom{n}{k}\binom{n}{j}\binom{n}{l_1} \cdots \binom{n}{l_s}} } } }H }}{{\left( {\sum\limits_{i = 0}^n {\omega _l B_l^n } } \right)^{s + 2} }},
\]
where
\begin{equation}\label{eq:001c}
H = \sum\limits_{p = 0}^s {\frac{{a_p^{[s]} B_{k + j + l_1  +  \cdots  + l_s  - 1 - s + p}^{(s + 2)(n - 1)} }}{\binom{(s + 2)(n - 1)}{k + j + l_1  +  \cdots  + l_s  - 1 - s + p}}}.
\end{equation}

In order to obtain geometric insights into the behavior of the derivatives of rational B\'{e}zier curves, Floater {\cite{Floater1992}} derived the following equations via de Casteljau algorithm
\begin{eqnarray}
 {\boldsymbol{c'}}(t) &= &n\frac{{\omega _{0,n - 1} \omega _{1,n - 1} }}{{\omega _{0,n}^2 }}\left( {{\boldsymbol{r}}_{1,n - 1}  - {\boldsymbol{r}}_{0,n - 1} } \right) \\
 {\boldsymbol{c''}}(t) &=& n\frac{{\omega _{2,n - 2} }}{{\omega _{0,n}^3 }}\left( {2n\omega _{0,n - 1}^2  - (n - 1)\omega _{0,n - 2} \omega _{0,n}  - 2\omega _{0,n - 1} \omega _{0,n} } \right)\left( {{\boldsymbol{r}}_{2,n - 2}  - {\boldsymbol{r}}_{1,n - 2} } \right) \nonumber\\
 {\rm{          }}& -& n\frac{{\omega _{0,n - 2} }}{{\omega _{0,n}^3 }}\left( {2n\omega _{1,n - 1}^2  - (n - 1)\omega _{2,n - 2} \omega _{0,n}  - 2\omega _{1,n - 1} \omega _{0,n} } \right)\left( {{\boldsymbol{r}}_{1,n - 2}  - {\boldsymbol{r}}_{0,n - 2} } \right) \nonumber  \label{eq:001a} \\
\end{eqnarray}

In 2009, Lin \cite{Lin2009} studied  expressions of the endpoint of the rational B\'{e}ziers curve based on equation \eqref{eq:001b} and described that the exact derivative for any parameter $t$ can be obtained by applying the de Casteljau subdivision algorithm. The key formula  in \cite{Lin2009} is
\[
{\boldsymbol{r}}^{(k)} (0) = \sum\limits_{i = 1}^k {\left( {A + \sum\limits_{j = 1}^{k - 1} {I_i^j } } \right)\left( {{\boldsymbol{r}}_i  - {\boldsymbol{r}}_0 } \right)},
\]
where
\begin{equation}\label{eq:001d}
  A = \frac{{n!}}{{(n - k)!}}( - 1)^{k - i} \binom{k}{i}\frac{\omega_i}{\omega_0}.
\end{equation}

The above equations \eqref{eq:001b}, \eqref{eq:001a} and \eqref{eq:001d} work well when the degree $n$ of a rational B\'{e}zier curve greater than the order $k$ of the derivative, but the reverse is not true. For example,  when $n=1$ and  $k=2$, we have the following results:
\begin{itemize}
 \item By equation \eqref{eq:001c}, we find that the denominator of $H$ is zero.
  \item By equation \eqref{eq:001a}, we obtain
\[
{\boldsymbol{c''}}(t) = \frac{{\omega _{2, - 1} }}{{\omega _{0,1}^3 }}\left( {2\omega _{0,0}^2  - 2\omega _{0,0} \omega _{0,1} } \right)\left( {{\boldsymbol{r}}_{2, - 1}  - {\boldsymbol{r}}_{1, - 1} } \right) - \frac{{\omega _{0, - 1} }}{{\omega _{0,1}^3 }}\left( {2\omega _{1,0}^2  - 2\omega _{1,0} \omega _{0,1} } \right)\left( {{\boldsymbol{r}}_{1, - 1}  - {\boldsymbol{r}}_{0, - 1} } \right),
\]
where the expressions ${\omega _{2, - 1} }, {\omega _{0, - 1} }, {{\boldsymbol{r}}_{2, - 1} }, {{\boldsymbol{r}}_{1, - 1} }$ and ${{\boldsymbol{r}}_{0, - 1} } $ have no specific meaning.
   \item By equation \eqref{eq:001d}, we get that $\frac{n!}{(n-k)!}=\frac{1}{(-1)!}$.
\end{itemize}

The  problems were discovered when we studied parametric curves that satisfy the $C^n$  continuity conditions \cite{Shi2020}\cite{Fang1999}\cite{Piegl1997}. In this paper, we provide a formula for computing the derivative of any order of a rational B\'{e}zier curve of arbitrary degree based on recursive algorithms. It can also be used to perform symbolic operations \cite{Piegl1997}.  The properties of the endpoints and the bounds for the derivatives also be considered.

The rest of this paper is organized as follows. In Section 2, we review the definition  of rational B\'{e}zier curves and some preliminary results  that will be needed later. In Section 3, we present our main results. As an example, we show the third-order derivative formula often used in engineering in Section 4. Finally, concludes the paper.

\section{Preliminary}
\begin{Lemma}
  $\sum\limits_{i = 0}^k \left( { - 1} \right)^{k - i}{\binom{k}{i} } {\rm{ }}= 0\ (k>0)$.
\end{Lemma}
\begin{Proof}
Since for any constant $a$, we have
\[
\sum\limits_{i = 0}^k {\left( { - 1} \right)^{k - i} }\binom{k}{i}a=\left( {E - I} \right)^k a {\rm{ = }}(a-a)^k=0,
\]
where $E$ is the shift operator and $I$ is the identity operator. This completes the proof.
\end{Proof}
\begin{Lemma}
   The sum of the coefficients of the control points $\boldsymbol{p}_{i+j},\ (i=0,...,k)$ in equation \eqref{eq:003} is zero.
\end{Lemma}
\begin{Proof}
Expanding $\Delta^k\boldsymbol{p}_j$ in equation \eqref{eq:003}, it yields
\[
\Delta ^k {\boldsymbol{p}}_j  = \sum\limits_{i = 0}^k {\left( { - 1} \right)^{k - i} } \binom{k}{i}{\boldsymbol{p}}_{i + j} ,
\]
and using lemma 1, we can obtain the result.
\end{Proof}
\begin{Lemma}\cite{Farouki1988}
Let ${\boldsymbol{q}}(t) $ be another B\'{e}zier curve of degree $m$ with  control points$\{{\boldsymbol{q}}_i\}_{i=0}^{m}$, then the product of ${\boldsymbol{q}}(t) $ and  ${\boldsymbol{p}}(t)$ \eqref{eq:002} can be written as
      \begin{eqnarray}\label{eq:004}
{\boldsymbol{p}}(t)\boldsymbol{q}(t)=\sum_{k=0}^{m+n}\sum\limits_{j=max(0,k-n)}^{min(m,k)}\frac{{m \choose j}{n \choose k-j}}{{m+n \choose k}}{\boldsymbol{p}}_{k-j}{\boldsymbol{q}}_{j}B_{k}^{m+n}(t).
\end{eqnarray}
\end{Lemma}
In addition, we set
\begin{equation}\label{eq:005}
  \prod\limits_{k = l}^m {x_k }  = \left\{ {\begin{array}{*{20}c}
   {x_l x_{l + 1}  \cdots x_m } & {l \le m},  \\
   1 & {otherwise}.  \\
\end{array}} \right.
\end{equation}

\section{Main results}
\subsection{Derivatives of rational B\'{e}zier curves}
\begin{Theorem}
  The $k$th derivative of a rational B\'{e}zier curve can be represented as
  \begin{equation}\label{eq:006}
    \boldsymbol {r}^{(k)}({t}) =\frac{\prod\limits_{j=0}^{k-1}{\left(2^j n \right)}\sum\limits_{i=0}^{2^k n}{\boldsymbol{\hat{P}}_{i,n}^{\left[ k \right]}}B_{i}^{2^k n}\left( t \right)}{\sum\limits_{i=0}^{2^k n}{\omega _{i,n}^{\left[ k \right]}B_{i}^{2^k n}\left( t \right)}},
  \end{equation}
where
\begin{equation}\label{eq:007}
\omega_{l,n}^{[0]}=\omega_l,\  \boldsymbol{P}_{l,n}^{[0]}=\boldsymbol{\hat{P}}_{l,n}^{[0]}=\omega_{l} \boldsymbol{r}_l,\ \ (l=0,...,n)
\end{equation}
\begin{gather}\label{eq:008}
  \omega _{i,n}^{\left[ k \right]}=\sum_{j=\max \left( 0,i-2^{k-1}n \right)}^{\min \left( i,2^{k-1}n \right)}{\frac{{
	2^{k-1}n\choose j
}{ 	2^{k-1}n\choose 	i-j
}}{{
	2^kn\choose 	i}}}\omega _{j,n}^{\left[ k-1 \right]}\omega _{i-j,n}^{\left[ k-1 \right]}, \\
(i=0,...,2^kn).  \nonumber
\end{gather}
\begin{gather}\label{eq:010}
  \boldsymbol{P}_{j,n}^{\left[ k \right]}=\sum_{h=\max \left( 0,j-2^{k-1}n \right)}^{\min \left( 2^{k-1}n-1,j \right)}{\frac{{
	2^{k-1}n-1\choose 	h
} {
	2^{k-1}n\choose 	j-h
}}{{
	2^kn-1\choose 	j
}}\left( \Delta \boldsymbol{\hat{P}}_{h,n}^{\left[ k-1 \right]}\omega _{j-h,n}^{\left[ k-1 \right]}-\Delta \omega _{h,n}^{\left[ k-1 \right]}\boldsymbol{\hat{P}}_{j-h,n}^{\left[ k-1 \right]} \right)}. \nonumber\\
  (j=0,...,2^kn-1)
\end{gather}
and
\begin{gather} \label{eq:009}
\boldsymbol{\hat{P}}_{i,n}^{\left[ k \right]}=\sum_{j=\max \left( 0,i-1 \right)}^{\min \left(2^{k}n-1,i \right)}{\frac{{
2^{k}n-1 \choose	j
} }{{
2^k n \choose	i
}}\boldsymbol{P}_{j,n}^{\left[ k \right]}}, \\
(i=0,...,2^kn).  \nonumber
\end{gather}
\end{Theorem}
\begin{Proof}
  The proof is by induction on $k$. It is  true for $k=0$ and $k=1$  based on equations\eqref{eq:003}, \eqref{eq:004}, \eqref{eq:005}, \eqref{eq:007}, \eqref{eq:008}, \eqref{eq:009} and \eqref{eq:010}.

Assuming it is valid for $k > 1$, then letting
$$
\boldsymbol{\hat{P}}^{\left[ k \right]}\left( t \right) =\sum_{i=0}^{2^kn}{\boldsymbol{\hat{P}}_{i,n}^{\left[ k \right]}}B_{i}^{2^kn}\left( t \right),
$$
and $$
\omega ^{\left[ k \right]}\left( t \right) =\sum_{i=0}^{2^kn}{\omega _{i,n}^{\left[ k \right]}B_{i}^{2^kn}\left( t \right)},
$$
we have
\begin{small}
\allowdisplaybreaks[4]
\begin{align*}\label{aa}
 & \boldsymbol{r}^{\left( k+1 \right)}\left( t \right) \\
=& \frac{\prod\limits_{j=0}^{k-1}{\left( 2^jn \right)}\left[ \left( \boldsymbol{\hat{P}}^{\left[ k \right]}\left( t \right) \right) '\omega ^{\left[ k \right]}\left( t \right) -\boldsymbol{\hat{P}}^{\left[ k \right]}\left( t \right) \left( \omega ^{\left[ k \right]}\left( t \right) \right) ' \right]}{\left( \omega ^{\left[ k \right]}\left( t \right) \right) ^2} \\
=&\frac{\prod\limits_{j=0}^k{( 2^jn )}\left[ \left( \sum\limits_{i=0}^{2^kn-1}{\varDelta \boldsymbol{\hat{P}}_{i,n}^{\left[ k \right]}}B_{i}^{2^kn-1}\left( t \right) \right) \omega ^{\left[ k \right]}\left( t \right) -\boldsymbol{\hat{P}}^{\left[ k \right]}\left( t \right) \left( \sum\limits_{i=0}^{2^kn-1}{\varDelta \omega _{i,n}^{\left[ k \right]}B_{i}^{2^kn-1}\left( t \right)} \right) \right]}{\left( \sum\limits_{i=0}^{2^kn}{\omega _{i,n}^{\left[ k \right]}B_{i}^{2^kn}\left( t \right)} \right) ^2}\\
=&\frac{\prod\limits_{j=0}^k{\left( 2^jn \right)}\sum\limits_{i=0}^{2^{k+1}n}{\left[ \sum\limits_{j=\max \left( 0,i-2^kn \right)}^{\min \left( 2^kn-1,i \right)}{\frac{{2^kn-1 \choose
	j}{	2^kn \choose i-j}}{{2^{k+1}n-1\choose i}}\left( \varDelta \boldsymbol{\hat{P}}_{j,n}^{\left[ k \right]}\omega _{i-j}^{\left[ k \right]}-\varDelta \omega _{j,n}^{\left[ k \right]}\boldsymbol{\hat{P}}_{i-j}^{\left[ k \right]} \right)} \right] B_{i}^{2^{k+1}n}\left( t \right)}}{ \sum\limits_{i=0}^{2^{k+1}n}{\left( \sum\limits_{j=\max \left( 0,i-2^kn \right)}^{\min \left( i,2^kn \right)}{\frac{{2^kn \choose j}{2^kn \choose i-j} }{{2^{k+1}n \choose i}}}\omega _{i,n}^{\left[ k \right]}\omega _{i-j,n}^{\left[ k \right]} \right) B_{i}^{2^{k+1}n}\left( t \right)} }\\
=&\frac{\prod\limits_{j=0}^k{\left( 2^jn \right)}\sum\limits_{i=0}^{2^{k+1}n}{\boldsymbol{\hat{P}}_{i,n}^{\left[ k+1 \right]}}B_{i}^{2^{k+1}n}\left( t \right)}{\sum\limits_{i=0}^{2^{k+1}n}{\omega _{i,n}^{\left[ k+1 \right]}B_{i}^{2^{k+1}n}\left( t \right)}},
\end{align*}
\end{small}
showing that \eqref{eq:006} holds for $k+1$. This completes the proof.
\end{Proof}

Applying Leibniz rule, lemma 1 and 2, $\boldsymbol{\hat{P}}_{i,n}^{\left[ k \right]}$ in equation \eqref{eq:010} has the following two properties.

\begin{Property}
$\boldsymbol{\hat{P}}_{i,n}^{\left[ k \right]}$ can be represented as a linear combination of $\boldsymbol{r}_{j}\ (j=0,...,n)$ in equation \eqref{eq:001} and the corresponding coefficients sum of $\boldsymbol{r}_{j}$ is zero.
\end{Property}
\begin{Property}
If we write $\boldsymbol{\hat{P}}_{j,n}^{\left[ k \right]}\left( \{\omega _i\},\{\boldsymbol{r}_i\} \right): =\boldsymbol{\hat{P}}_{j,n}^{\left[ k \right]}$ and replace $\omega _i$ and $\boldsymbol{r}_i$ by  $\omega _{n-i}$ and $\boldsymbol{r}_{n-i}$, respectively, we have
$$
\boldsymbol{\hat{P}}_{j,n}^{\left[ k \right]}\left( \{\omega _i\},\{\boldsymbol{r}_i\}\right) =\left( -1 \right) ^k\boldsymbol{\hat{P}}_{2^kn-j,n}^{\left[ k \right]}\left(  \{\omega _{n-i}\},\{\boldsymbol{r}_{n-i}\} \right),
$$
$$(j=0,...,2^{k-1}n-1, i=0,...,min\{j+1,n\}).$$
\end{Property}
\textbf{Notes:} 1) The coefficients of $\boldsymbol{r}_{i} $ sum to zero  indicating that $\boldsymbol{\hat{P}}_{j,n}^{\left[ k \right]}$ and can be expressed  in various forms such as $\varDelta \boldsymbol{r}_i $ {\cite{wang1995}} or $\boldsymbol{r}_i-\boldsymbol{r}_0$ \cite{Lin2009}.

2) By property 2, we can simplify the process of finding  $\boldsymbol{\hat{P}}_{2^kn-j,n}^{\left[ k \right]}$, $(j=0,...,2^{k-1}n-1)$. 

For more details, please refer to  example 1 and example 2 in this paper.

\subsection{Calculation of corners}
In  Computer Aided Geometric Design (CAGD), the  smoothness of joining two B\'{e}zier curves requires knowledge of the derivative formula at their merging corners, for which we have the following conclusions.

\begin{Theorem}
  The derivative of the rational B\'{e}zier curve  at  $t=0$ is
  \begin{eqnarray}
    {\boldsymbol{r}}^{(k)} (0) &=& \frac{{2^{\frac{1}{2}k(k - 1)} n^k }}{{\left( {\omega _0 } \right)^{2^k } }}{\boldsymbol{\hat{P}}}_{0,n}^{[k]}\left( {\{\omega _{i}\} ,\{{\boldsymbol{r}}_{i} }\} \right)  \ \  \ \ \ \  \ \ \ \ \ \ \ \ \ \  (k \ge 0) \label{eq:011} \\
     &=& \frac{{2^{\frac{1}{2}k(k - 1)} n^k }}{{\left( {\omega _0 } \right)^{2^{k - 1} } }}\left( {{\boldsymbol{\hat P}}_1^{\left[ {k - 1} \right]}\left( {\{\omega _{i}\} ,\{{\boldsymbol{r}}_{i} }\} \right)  - \frac{{\omega _1 }}{{\omega _0 }}{\boldsymbol{ P}}_0^{\left[ {k - 1} \right]}\left( {\{\omega _{i}\} ,\{{\boldsymbol{r}}_{i} }\} \right) } \right)\nonumber \\
     & &  \ \ \ \  \ \ \ \ \  \ \ \ \ \  \ \ \ \ \ \  \ \ \ \ \ \ \  \  \ \ \ \ \ \ \ \ \ \ \ \ \  \ \ \ \ \ \ \ \ \  (k \ge 1)    \label{eq:012}
  \end{eqnarray}
and at  $t=1$ can be expressed as
\begin{small}
\begin{eqnarray}
 {\bf{r}}^{(k)} (1) &=&  (-1)^k \frac{{2^{\frac{1}{2}k(k - 1)} n^k }}{{\left( {\omega _n } \right)^{2^k } }}{\boldsymbol{\hat P}}_{0,n}^{[k]} \left( {\{\omega _{n - i}\} ,\{{\boldsymbol{r}}_{n - i} }\} \right)\;\;\;\;\;\;\;\;\;\;\;\;\;\;\;\;(k \ge 0) \label{eq:013}\\
 {\rm{        }} &=& (-1)^k\frac{{2^{\frac{1}{2}k(k - 1)} n^k }}{{\left( {\omega _n } \right)^{2^{k - 1} } }}\left( {{\boldsymbol{\hat P}}_1^{\left[ {k - 1} \right]} \left( {\{\omega _{n - i}\} ,\{{\boldsymbol{r}}_{n - i} }\} \right) - \frac{{\omega _{n - 1} }}{{\omega _n }}{\boldsymbol{P}}_0^{\left[ {k - 1} \right]} \left( {\{\omega _{n - i}\} ,\{{\boldsymbol{r}}_{n - i} }\} \right)} \right) \nonumber\\
& & \;\;\;\;\;\;\;\;\;\;\;\;\;\;\;\;\;\;\;\;\;\;\;\;\;\;\;\;\;\;\;\;\;\;\;\;\;\;\;\;\;\;\;\;\;\;\;\;\;\;\;\;\;\;\;\;\;\;\;\;\;\;\;\;\;\;\;\;\;\;\;\;\;\;\;\ \ \ (k \ge 1) \label{eq:014}
 \end{eqnarray}
\end{small}
\end{Theorem}
\begin{Proof}
It is clear that
\begin{equation}\label{eq:015}
  \prod\limits_{j = 0}^{k - 1} {\left( {2^j n} \right)}  = 2^{{\textstyle{1 \over 2}}k(k - 1)} n^k .
\end{equation}
By equations \eqref{eq:007} and \eqref{eq:008}, we get
\begin{eqnarray}
  \omega _0^{\left[ k \right]}  &=& \sum\limits_{j = \max \left( {0,0 - 2^{k - 1} n} \right)}^{\min \left( {0,2^{k - 1} n} \right)} {\frac{{\binom{2^{k - 1} n}{j}\binom{2^{k - 1} n}{0 - j}}}{{\binom{2^k n}{0}}}} \omega _j^{\left[ {k - 1} \right]} \omega _{0 - j}^{\left[ {k - 1} \right]} \nonumber  \\
  &=&  \left( {\omega _0^{\left[ {k - 1} \right]} } \right)^2  = \left( {\omega _0^{\left[ {k - 2} \right]} } \right)^{2^2 }  \nonumber \\
   &=&   \cdots  \cdots   \nonumber \\
   &=& \left( {\omega _0 } \right)^{2^k },  \label{eq:016}
\end{eqnarray}
and based on equations \eqref{eq:009}and \eqref{eq:010}, it yields
\begin{eqnarray}
  {\boldsymbol{\hat P}}_0^{\left[ k \right]}  &=& {\boldsymbol{P}}_0^{\left[ k \right]}  \nonumber \\
    &=& {\boldsymbol{\hat P}}_1^{\left[ {k - 1} \right]} \omega _0^{\left[ {k - 1} \right]}  - \omega _1^{\left[ {k - 1} \right]} {\boldsymbol{\hat P}}_0^{\left[ {k - 1} \right]} {\rm{ }}. \label{eq:017}
\end{eqnarray}
Finally, substituting equations \eqref{eq:015}, \eqref{eq:016} and \eqref{eq:017} into  ${\boldsymbol{r}}^{(k)} (0)$, we can obtain \eqref{eq:011}  and \eqref{eq:012}.

Additionally, Based on Property 2 and Theorem 2, we have equations \eqref{eq:013} and \eqref{eq:014}. This completes the proof.
\end{Proof}


\subsection{Bounds for the derivatives}
Using equations \eqref{eq:004}, \eqref{eq:006} and  the following inequality \cite{Kuang1989}
  \begin{equation*}
\frac{{\sum {a_k c_k } }}
{{\sum {b_k c_k } }} \leqslant \mathop {\max }\limits_k \frac{{a_k }}
{{b_k }},
\end{equation*}
where $a_k \in\mathbb{R}$, $b_k>0 $, $c_k>0$ and $\ 1\leq k \leq n$, we  obtain an upper bound of the $k$th derivative of the rational B\'{e}zier curve.
\begin{Theorem}
\[
\left\| {{\boldsymbol{r}}^{\left( k \right)} \left( t \right)} \right\|_{l^p} \le\mathop {\max }\limits_i \frac{ \prod\limits_{j = 0}^{k - 1} {\left( {2^j n} \right)}{\left\| {\sum\limits_{j = \max (0,i - e)}^{\min (2^k n,i)} {{{2^k n}  \choose  j }{ e  \choose
   {i - j} } {\boldsymbol{\hat P}}_j^{\left[ k \right]} } } \right\|_{l^p}}}{{\sum\limits_{j = \max (0,i - e)}^{\min (2^k n,i)} {{{2^k n}  \choose  j }{ e  \choose
   {i - j} }\omega _j^{\left[ k \right]} } }},\ \ \ (i=0,...,2^kn+e),
\]
where $e$ is the number of degree elevation and $l^ p$ is the vector norm.
\end{Theorem}

\section{Example}
Here, we provide several examples to demonstrate the relevance of the earlier ideas.

\noindent \textbf{Example 1.}
For $k=1$, by Theorem 1, we  can obtain the following results:

When $n=1$,
$$
\boldsymbol{\hat{P}}_{0,1}^{\left[ 1 \right]}=\boldsymbol{\hat{P}}_{1,1}^{\left[ 1 \right]}=\boldsymbol{\hat{P}}_{2,1}^{\left[ 1 \right]}=-\omega _0\omega _1\boldsymbol{r}_0+\omega _0\omega _1\boldsymbol{r}_1.
$$

When $n=2$,
\begin{align*}
 \boldsymbol{\hat{P}}_{0,2}^{\left[ 1 \right]}=&-\omega _0\omega _1\boldsymbol{r}_0+\omega _0\omega _1\boldsymbol{r}_1, \\
 \boldsymbol{\hat{P}}_{1,2}^{\left[ 1 \right]}=&-\frac{\omega _0\left( 2\omega _1+\omega _2 \right)\boldsymbol{r}_0}{4}+\frac{\omega _0\omega _1\boldsymbol{r}_1}{2}+\frac{\omega _0\omega _2\boldsymbol{r}_2}{4}, \\
 \boldsymbol{\hat{P}}_{2,2}^{\left[ 1 \right]}=&-\frac{\omega _0\left( \omega _1+2\omega _2 \right) \boldsymbol{r}_0}{6}+\frac{\left( \omega _0-\omega _2 \right) \omega _1\boldsymbol{r}_1}{6}+\frac{\left( \omega _1+2\omega _0 \right) \omega _2\boldsymbol{r}_2}{6},\\
 \boldsymbol{\hat{P}}_{3,2}^{\left[ 1 \right]}=& -\boldsymbol{\hat{P}}_{1,2}^{\left[ 1 \right]}\left( \omega _{2-i},\boldsymbol{r}_{2-i} \right)\\
 =&-\frac{\omega _0\omega _2\boldsymbol{r}_0}{4}-\frac{\omega _1\omega _2\boldsymbol{r}_1}{2}+\frac{\left( \omega _0+2\omega _1 \right) \omega _2\boldsymbol{r}_2}{4},\\
 \boldsymbol{\hat{P}}_{4,2}^{\left[ 1 \right]}=&- \boldsymbol{\hat{P}}_{0,2}^{\left[ 1 \right]}\left( \omega _{2-i},\boldsymbol{r}_{2-i} \right)\\
 =&-\omega _1\omega _2\boldsymbol{r}_1+\omega _2\omega _1\boldsymbol{r}_2.
\end{align*}

When $n=3$,
\begin{align*}
  \boldsymbol{\hat{P}}_{0,3}^{\left[ 1 \right]}=&-\omega _0\omega _1\boldsymbol{r}_0+\omega _0\omega _1\boldsymbol{r}_1, \\
\boldsymbol{\hat{P}}_{1,3}^{\left[ 1 \right]}=&-\frac{\omega _0\left( \omega _1+\omega _2 \right) \boldsymbol{r}_0}{3}+\frac{\omega _0\omega _1\boldsymbol{r}_1}{3}+\frac{\omega _0\omega _2\boldsymbol{r}_2}{3}, \\
\boldsymbol{\hat{P}}_{2,3}^{\left[ 1 \right]}=&-\frac{\omega _0\left( \omega _1+4\omega _2+\omega _3 \right) \boldsymbol{r}_0}{15}+\frac{\omega _1\left( \omega _0-3\omega _2 \right) \boldsymbol{r}_1}{15}+\frac{\left( 4\omega _0+3\omega _1 \right) \omega _2\boldsymbol{r}_2}{15}+\frac{\omega _0\omega _3\boldsymbol{r}_3}{15},\\
\boldsymbol{\hat{P}}_{3,3}^{\left[ 1 \right]}=& \frac{\omega _0\left( \omega _2+\omega _3 \right) \boldsymbol{r}_0}{10}-\frac{\left( 3\omega _2+\omega _3 \right) \omega _1\boldsymbol{r}_1}{10}+\frac{\omega _2\left( 3\omega _1+\omega _0 \right) \boldsymbol{r}_2}{10}+\frac{\omega _3\left( \omega _1+\omega _0 \right) \boldsymbol{r}_3}{10}, \\
\boldsymbol{\hat{P}}_{4,3}^{\left[ 1 \right]}=&- \boldsymbol{\hat{P}}_{2,3}^{\left[ 1 \right]}\left( \omega _{3-i},\boldsymbol{r}_{3-i} \right)\\
=&-\frac{\omega _0\omega _3\boldsymbol{r}_0}{15}-\frac{\omega _1\left( 4\omega _3+3\omega _2 \right) \boldsymbol{r}_1}{15}+\frac{\omega _2\left( 3\omega _1-\omega _3\right) \boldsymbol{r}_2}{15}+\frac{\left( \omega _0+4\omega _1+\omega _2 \right) \omega _3\boldsymbol{r}_3}{15}, \\
\boldsymbol{\hat{P}}_{5,3}^{\left[ 1 \right]}=&- \boldsymbol{\hat{P}}_{1,3}^{\left[ 1 \right]}\left( \omega _{3-i},\boldsymbol{r}_{3-i} \right)\\
=&-\frac{\omega _1\omega _3\boldsymbol{r}_1}{3}-\frac{\omega _2\omega _3\boldsymbol{r}_2}{3}+\frac{\left( \omega _1+\omega _2 \right) \omega _3\boldsymbol{r}_3}{3}, \\
\boldsymbol{\hat{P}}_{6,3}^{\left[ 1 \right]}=&- \boldsymbol{\hat{P}}_{0,3}^{\left[ 1 \right]}\left( \omega _{3-i},\boldsymbol{r}_{3-i} \right)\\
=&-\omega _2\omega _3\boldsymbol{r}_2+\omega _2\omega _3\boldsymbol{r}_3.
\end{align*}

\noindent \textbf{Example 2.}
For $k=2$, we  can obtain the following results:

When $n=1$,
\begin{align*}
  \boldsymbol{\hat{P}}_{0,1}^{\left[ 2 \right]} =& -\omega _{0}^{2}\omega _1\left( \omega _0-\omega _1 \right) \boldsymbol{r}_0+\omega _{0}^{2}\omega _1\left( \omega _0-\omega _1 \right) \boldsymbol{r}_1, \\
  \boldsymbol{\hat{P}}_{1,1}^{\left[ 2 \right]} =& -\frac{\omega _0\omega _1\left( \omega _0-\omega _1 \right) \left( 3\omega _0+\omega _1 \right) \boldsymbol{r}_0}{4}+\frac{\omega _1\omega _0\left( \omega _0-\omega _1 \right) \left( 3\omega _0+\omega _1 \right) \boldsymbol{r}_1}{4}, \\
  \boldsymbol{\hat{P}}_{2,1}^{\left[ 2 \right]} =& -\frac{\omega _0\omega _1\left( \omega _0-\omega _1 \right) \left( \omega _0+\omega _1 \right) \boldsymbol{r}_0}{2}+\frac{\omega _0\omega _1\left( \omega _0-\omega _1 \right) \left( \omega _0+\omega _1 \right) \boldsymbol{r}_1}{2}, \\
  \boldsymbol{\hat{P}}_{3,1}^{\left[ 2 \right]}=&\boldsymbol{\hat{P}}_{1,1}^{\left[ 2 \right]}\left( \omega _{1-i},\boldsymbol{r}_{1-i} \right)\\
  =&-\frac{\omega _0\omega _1\left( \omega _0-\omega _1 \right) \left( \omega _0+3\omega _1 \right) \boldsymbol{r}_0}{4}+\frac{\omega _0\omega _1\left( \omega _0+3\omega _1 \right) \left( \omega _0-\omega _1 \right) \boldsymbol{r}_1}{4},\\
  \boldsymbol{\hat{P}}_{4,1}^{\left[ 2 \right]} =& \boldsymbol{\hat{P}}_{0,1}^{\left[ 2 \right]}\left( \omega _{1-i},\boldsymbol{r}_{1-i} \right) \\
  =& -\omega _0\omega _{1}^{2}\left( \omega _0-\omega _1 \right) \boldsymbol{r}_0+\omega _0\omega _{1}^{2}\left( \omega _0-\omega _1 \right) \boldsymbol{r}_1.
\end{align*}

When $n=2$,
\begin{align*}
  \boldsymbol{\hat{P}}_{0,2}^{\left[ 2 \right]} = &\frac{\omega _{0}^{3}\omega _2\boldsymbol{r}_2}{4}+\frac{\omega _{0}^{2}\left( \omega _0\omega _1-2\omega _{1}^{2} \right) \boldsymbol{r}_1}{2}-\frac{\omega _{0}^{2}\left( \left( 2\omega _1+\omega _2 \right) \omega _0-4\omega _{1}^{2} \right) \boldsymbol{r}_0}{4}, \\
  \boldsymbol{\hat{P}}_{1,2}^{\left[ 2 \right]} =& \frac{\omega _0\left( 3\omega _{0}^{2}\omega _2+\omega _0\omega _1\omega _2 \right) \boldsymbol{r}_2}{16}+\frac{\omega _0\left( 3\omega _{0}^{2}\omega _1-\omega _0\omega _1\left( 4\omega _1+3\omega _2 \right) -4\omega _{1}^{3} \right) \boldsymbol{r}_1}{16}
 \\
   & -\frac{\omega _0\left( 3\left( \omega _1+\omega _2 \right) \omega _{0}^{2}-\omega _1\left( 4\omega _1+2\omega _2 \right) \omega _0-4\omega _{1}^{3} \right) \boldsymbol{r}_0}{16},
 \\
\boldsymbol{\hat{P}}_{2,2}^{\left[ 2 \right]}=&\frac{\omega _2\omega _{0}^{2}\left( 6\omega _0+9\omega _1-\omega _2 \right) \boldsymbol{r}_2}{56}-\frac{\left( 11\omega _0\omega _1\omega _2-3\omega _1\omega _{0}^{2}+4\omega _{1}^{2}\left( 3\omega _1+2\omega _2 \right) \right) \omega _0\boldsymbol{r}_1}{56}
\\
&-\frac{\omega _0\left( 3\left( \omega _1+2\omega _2 \right) \omega _{0}^{2}-\left( 2\omega _1+\omega _2 \right) \omega _2\omega _0-4\left( 3\omega _1+2\omega _2 \right) \omega _{1}^{2} \right) \boldsymbol{r}_0}{56},
\\
\boldsymbol{\hat{P}}_{3,2}^{\left[ 2 \right]}=&\frac{\omega _0\left( 5\omega _{0}^{2}+21\omega _0\omega _1-2\omega _0\omega _2+8\omega _{1}^{2}-4\omega _1\omega _2 \right) \omega _2\boldsymbol{r}_2}{112}
\\
 &-\frac{\left( 12\omega _{1}^{3}+4\left( 8\omega _2-\omega _0 \right) \omega _1-\omega _{0}^{2}+15\omega _0\omega _2+2\omega _{2}^{2} \right) \omega _0\omega _1\boldsymbol{r}_1}{112}
\\
&-\frac{\omega _0\left( -12\omega _{1}^{3}-4\left( 6\omega _2-\omega _0 \right) \omega _{1}^{2}-\left( 6\omega _{2}^{2}-6\omega _0\omega _2-\omega _{0}^{2} \right) \omega _1-2\omega _0\omega _{2}^{2}+5\omega _{0}^{2}\omega _2 \right) \boldsymbol{r}_0}{112},
\\
\boldsymbol{\hat{P}}_{4,2}^{\left[ 2 \right]}=&-\frac{\omega _0\left( -3\omega _{2}^{3}-38\omega _{2}^{2}\omega _1+\left( 3\omega _{0}^{2}+20\omega _0\omega _1-48\omega _{1}^{2} \right) \omega _2+4\omega _{1}^{2}\left( \omega _0-2\omega _1 \right) \right) \boldsymbol{r}_0}{280}
\\
&+\frac{\left( \left( -4\omega _0-4\omega _2 \right) \omega _{1}^{2}+2\left( \omega _{0}^{2}-24\omega _0\omega _2+\omega _{2}^{2} \right) \omega _1-9\omega _2\omega _0\left( \omega _0+\omega _2 \right) \right) \omega _1\boldsymbol{r}_1}{140}
\\
&+\frac{\left( 8\omega _{1}^{3}+\left( 48\omega _0-4\omega _2 \right) \omega _{1}^{2}+\left( 38\omega _0-20\omega _2 \right) \omega _0\omega _1+3\omega _{0}^{3}-3\omega _0\omega _{2}^{2} \right) \omega _2\boldsymbol{r}_2}{280},
\\
\boldsymbol{\hat{P}}_{5,2}^{\left[ 2 \right]}=&\boldsymbol{\hat{P}}_{3,2}^{\left[ 2 \right]}\left( \omega _{3-i},\boldsymbol{r}_{3-i} \right)\\=&\frac{\omega _0\left( 5\omega _{2}^{2}+21\omega _1\omega _2-2\omega _0\omega _2+8\omega _{1}^{2}-4\omega _0\omega _1 \right) \omega _2\boldsymbol{r}_0}{112}
\\
&-\frac{\left( 12\omega _{1}^{2}+4\left( 8\omega _0-\omega _2 \right) \omega _1+2\omega _{0}^{2}+15\omega _0\omega _2-\omega _{2}^{2} \right) \omega _2\omega _1\boldsymbol{r}_1}{112}
\\
&-\frac{\left( -12\omega _{1}^{3}-4\left( 6\omega _0-\omega _2 \right) \omega _{1}^{2}-\left( 6\omega _{0}^{2}-6\omega _0\omega _2-\omega _{2}^{2} \right) \omega _1-2\omega _{0}^{2}\omega _2+5\omega _0\omega _{2}^{2} \right) \omega _2\boldsymbol{r}_2}{112},
\\
\boldsymbol{\hat{P}}_{6,2}^{\left[ 2 \right]}=&\boldsymbol{\hat{P}}_{2,2}^{\left[ 2 \right]}\left( \omega _{3-i},\boldsymbol{r}_{3-i} \right)\\
=&\frac{\omega _0\omega _{2}^{2}\left( 6\omega _2+9\omega _1-\omega _0 \right) \boldsymbol{r}_0}{56}-\frac{\left( 11\omega _0\omega _1\omega _2-3\omega _1\omega _{2}^{2}+4\omega _{1}^{2}\left( 2\omega _0+3\omega _1 \right) \right) \omega _2\boldsymbol{r}_1}{56}
\\
&-\frac{\left( 3\left( 2\omega _0+\omega _1 \right) \omega _{2}^{2}-\left( \omega _0+2\omega _1 \right) \omega _0\omega _2-4\omega _{1}^{2}\left( 2\omega _0+3\omega _1 \right) \right) \omega _2\boldsymbol{r}_2}{56},
\\
 \boldsymbol{\hat{P}}_{7,2}^{\left[ 2 \right]}=&\boldsymbol{\hat{P}}_{1,2}^{\left[ 2 \right]}\left( \omega _{3-i},\boldsymbol{r}_{3-i} \right)\\
 =&\frac{\left( \omega _0\omega _1\omega _2+3\omega _0\omega _{2}^{2} \right) \omega _2\boldsymbol{r}_0}{16}+\frac{\left( 3\omega _1\omega _{2}^{2}-\left( 4\omega _1+3\omega _0 \right) \omega _1\omega _2-4\omega _{1}^{3} \right) \omega _2\boldsymbol{r}_1}{16}\\
&-\frac{\left( \left( 3\omega _0+3\omega _1 \right) \omega _{2}^{2}-\left( 2\omega _0+4\omega _1 \right) \omega _1\omega _2-4\omega _{1}^{3} \right) \omega _2\boldsymbol{r}_2}{16},
 \\
 \boldsymbol{\hat{P}}_{8,2}^{\left[ 2 \right]}= & \boldsymbol{\hat{P}}_{0,2}^{\left[ 2 \right]}\left( \omega _{3-i},\boldsymbol{r}_{3-i} \right)
 \\=&\frac{\omega _{2}^{3}\omega _0\boldsymbol{r}_0}{4}+\frac{\left( \omega _1\omega _2-2\omega _{1}^{2} \right) \omega _{2}^{2}\boldsymbol{r}_1}{2}-\frac{\left( \left( \omega _0+2\omega _1 \right) \omega _2-4\omega _{1}^{2} \right) \omega _{2}^{2}\boldsymbol{r}_2}{4}.
\end{align*}

\noindent \textbf{Example 3.}

  1) When $n=1$, we have
 $${\boldsymbol{r'''}}(0) = \frac{{6\omega _1 (\omega _0  - \omega _1 )^2 \left( {{\boldsymbol{r}}_1  - {\boldsymbol{r}}_0 } \right)}}{{\omega _0^3 }},$$
 and
 $${\boldsymbol{r'''}}(1) = \frac{{6\omega _0 ( - \omega _1  + \omega _0 )^2 \left( {{\boldsymbol{r}}_1  - {\boldsymbol{r}}_0 } \right)}}{{\omega _1^3 }}.$$

  2) When $n=2$, we have
 $${\boldsymbol{r'''}}(0) = \frac{{12(\omega _0  - 2\omega _1 )((\omega _1  + \omega _2 )\omega _0  - 2\omega _1^2 )}}{{\omega _0^3 }}\left( {{\boldsymbol{r}}_1  - {\boldsymbol{r}}_0 } \right) - \frac{{12\omega _2 (\omega _1  - \omega _0 )}}{{\omega _0^2 }}\left( {{\boldsymbol{r}}_2  - {\boldsymbol{r}}_1 } \right),$$
 and
 \[{\boldsymbol{r'''}}(1) = \frac{{12\omega _0 (\omega _2  - \omega _1 )}}{{\omega _2^2 }}\left( {{\boldsymbol{r}}_1  - {\boldsymbol{r}}_0 } \right) - \frac{{12((\omega _0  + \omega _1 )\omega _2  - 2\omega _1^2 )(2\omega _1  - \omega _2 )}}{{\omega _2^3 }}\left( {{\boldsymbol{r}}_2  - {\boldsymbol{r}}_1 } \right).\]

  3) When $n\geq3$, we have
\[
\begin{array}{l}
 {\boldsymbol{r'''}}(0) = \frac{{n((n^2 \omega _3  + (6\omega _2  - 3\omega _3 )n + 6\omega _1  - 6\omega _2  + 2\omega _3 )\omega _0^2  - 6n\omega _1 (n\omega _2  + 2\omega _1  - \omega _2 )\omega _0  + 6\omega _1^3 n^2 )}}{{\omega _0^3 }}\left( {{\boldsymbol{r}}_1  - {\boldsymbol{r}}_0 } \right) \\
  + \frac{{(n - 1)((n\omega _3  + 6\omega _2  - 2\omega _3 )\omega _0  - 3n\omega _1 \omega _2 )n}}{{\omega _0^2 }}\left( {{\boldsymbol{r}}_2  - {\boldsymbol{r}}_1 } \right) + \frac{{(n^2  - 3n + 2)n\omega _3 }}{{\omega _0 }}\left( {{\boldsymbol{r}}_3  - {\boldsymbol{r}}_2 } \right). \\
 \end{array}
\]
 and
  \[
\begin{array}{l}
 {\boldsymbol{r'''}}(1) \\
  = \frac{{n(n - 1)((n\omega _{n - 3}  + 6\omega _{n - 2}  - 2\omega _{n - 3} )\omega _n  - 3n\omega _{n - 1} \omega _{n - 2} )}}{{\omega _n^2 }}\left( {{\boldsymbol{r}}_{n - 1}  - {\boldsymbol{r}}_{n - 2} } \right) + \frac{{(n^2  - 3n + 2)n\omega _{n - 3} }}{{\omega _n }}\left( {{\boldsymbol{r}}_{n - 2}  - {\boldsymbol{r}}_{n - 3} } \right) +  \\
 \frac{{((n^2 \omega _{n - 3}  + (6\omega _{n - 2}  - 3\omega _{n - 3} )n + 6\omega _{n - 1}  - 6\omega _{n - 2}  + 2\omega _{n - 3} )\omega _n^2  - 6n\omega _{n - 1} (n\omega _{n - 2}  + 2\omega _{n - 1}  - \omega _{n - 2} )\omega _n  + 6\omega _{n - 1}^3 n^2 )}}{{\omega _n^3 }}\left( {{\boldsymbol{r}}_n  - {\boldsymbol{r}}_{n - 1} } \right). \\
 \end{array}
\]
\section{Conclusion}
  A complete and correct derivation method of rational B\'{e}zier curves is proposed, which can be extended to rational B\'{e}zier surfaces.
\section{Acknowledgements}
This paper is supported by  Natural Science Foundation of Shaanxi Province (No. 2013JM1004).
\section{Declaration of competing interest}
The authors declare that they have no known competing financial interests or personal relationships that could have appeared to influence the work reported in this paper.


\bibliography{mybibfile}

\end{document}